\documentclass[useAMS,usenatbib,letter]{mn2e}
\usepackage{graphicx}

\title[Circumstellar and Circumbinary Disks in Eccentric Stellar Binaries]
{Circumstellar and Circumbinary Disks in Eccentric Stellar Binaries}
\author[B. Pichardo L.S. Sparke L.A. Aguilar]{Barbara 
Pichardo$^{1,2}$\thanks{E-mail:
barbara@pa.uky.edu (BP); sparke@astro.wisc.edu (LSS); 
aguilar@astrosen.unam.mx (LAA)},
Linda S. Sparke$^{1}$\footnotemark[1], Luis A. Aguilar$^{3}$\footnotemark[1]\\
$^{1}$Department of Astronomy, 475 North Charter Street, University of 
Wisconsin-Madison,
Madison WI 53706-1582, USA\\
$^{2}$Department of Physics and Astronomy, University of Kentucky, Lexington, KY 40506-0055, USA\\
$^{3}$Observatorio Astron\'omico Nacional, IAUNAM Ensenada, Apdo. postal 877, 
22800
Ensenada, M\'exico}
\begin{document}

\date{Accepted . Received ; in original form }

\pagerange{\pageref{firstpage}--\pageref{lastpage}} \pubyear{}

\maketitle

\label{firstpage}

\begin{abstract} We explore test particle orbits in the orbital plane
of eccentric stellar binary systems, searching for ``invariant
loops'': closed curves that change shape periodically as a function of
binary orbital phase as the test particles in them move under the
stars' gravity. Stable invariant loops play the same role in this
periodically-varying potential as stable periodic orbits do in
stationary potentials; in particular, when dissipation is weak, gas
will most likely follow the non-intersecting loops, while nearby
particle orbits librate around them. We use this method to set bounds
on the sizes of disks around the stars, and on the gap between those
and the inner edge of a possible circumbinary disk.  Gas dynamics may
impose further restrictions, but our study sets upper bounds for the
size of circumstellar disks, and a lower bound for the inner radius of
a circumbinary disk.  We find that circumstellar disks are sharply
reduced as the binary's eccentricity grows.  For the disk around the
secondary star, the tidal (Jacobi) radius calculated for circular
orbits at the periastron radius, gives a good estimate of the maximum
size. Disks change in size and shape only marginally with the binary
phase,  with no strong  preference to increase or decrease at any
particular phase. The circumstellar disks in particular can be  quite
asymmetric.  We compare our results with other numerical and
theoretical results and with observations of the $\alpha$ Centauri and
L1551 systems, finding very good agreement. The calculated changes in
the shapes and crowding of the circumstellar orbits can be used to
predict how the disk luminosity and mass inflow should vary with
binary phase.

\end{abstract}


\section{Introduction}\label{Intro}

Most low-mass main--sequence stars are members of binary or multiple
systems (Duquennoy \& Mayor 1991, Fischer \& Marcy 1992).  The binary
frequency of pre-main-sequence stars is also high, and probably higher
than for main-sequence stars (Mathieu, Walter \& Myers 1989; Simon et
al. 1992; Ghez, Neugebauer \& Matthews 1993; Leinert 1993; Reipurth \&
Zinnecker 1993). In the last decade, interest in these systems has
increased significantly with the discovery that many T-Tauri and other
pre-main-sequence binary stars, possess both circumstellar and
circumbinary disks (for a review see Mathieu 1994). Observations of
binary systems suggest the existence of disk material around one or
both stars, as is inferred from observations of excess radiation at
infrared to millimeter wavelengths, polarization, and both Balmer
and forbidden emission lines (Mathieu 2000).  Some extrasolar planets
have been found to orbit stars that have a stellar companion, e.g., 16
Cygni B, $\tau$ Bootis, and 55 $\rho$ Cancri (Butler et al. 1997;
Cochran et al. 1997), confirming that planets can form in binary star
systems.  Thus the study of stellar disks in binary systems, as well
as the possibility of stable orbits, is a key element for better
understanding stellar and planetary formation.

It is currently believed that multiple stellar systems result from
fragmentation, which produces mainly eccentric binaries (Bonnell \&
Bastien 1992, Bate 1997; Bate \& Bonnell 1997), in particular wide
stellar systems with separations $\ge$ 10 $AU$ (Bate, Bonnell \& Bromm
2002). Although main-sequence binary systems typically have eccentric
orbits (Duquennoy \& Mayor 1991), most theoretical studies have
focused on binaries in near-circular orbits.  Extensive and very good
theoretical work has been done here (Lubow \& Shu 1975; Paczy\'nski
1977; Rudak \& Paczy\'nski 1981; Papaloizou \& Pringle 1977; Bonnell
\& Bastien 1992, Bate 1997; Bate \& Bonnell 1997).

Only a few observed pre-main sequence binaries have known orbital
elements (Pogodin et al. 2004; Schaefer et al. 2003; Masciadri \& Raga
2002) and information on the accompanying disks is even rarer (Kastner
2004; Jensen et al. 2004; Nielbock 2003; Liu 2003; Rodriguez et
al. 1998, and references therein). From the observational point of
view, measuring the size of a circumstellar disk is quite difficult,
since these are typically $< 100\ AU$  across.  Modelers also have a
hard task, since dynamical effects like resonances can introduce very
fine and complicated structure. In circular binaries, where the
gravitational potential is fixed in a uniformly-rotating frame, and an
energy-like integral is conserved along particle orbits, the position
and strength of resonances can be calculated. In the interesting
eccentric case, we lack a conserved integral, and it is not clear how
to proceed. One of the most important studies to date is by Artymowicz
\& Lubow (1994), who compute analytically the position of orbital
resonances (between 1:4 and 1:3) in the circumstellar disks, and
approximate the sizes of disks by computing the radius where resonant
and viscous torques balance each other. Their results, however, depend
strongly on the badly-constrained viscosity parameter.

The complexity of the problem has prompted some researchers to use
direct $3$-body simulations to study the possible existence of stable
orbits for planets around eccentric orbits (David et al. 2003 and
references therein). This approach, however, is computationally
expensive when accuracy is required. Others have used hydrodynamic
simulations (Foulkes et al. 2004; Guerrero, Garc\'\i a-Berro, Isern
2004; Lanzafame 2003 and references therein). Unfortunately these
simulations depend sensitively on the unknown viscosity as well, 
and are quite expensive computationally.

In this work we have opted for a simpler approach, analogous to using
the structure of periodic orbits in a circular binary, to predict the
gas flow. The path followed by a gas parcel in a stable disk around a
star  must not intersect itself, or the path of a neighboring parcel
(in the case of planets, the paths may cross and we consider this in
the application to a planetary system). Using a test particle method,
we probe the orbital structure of binaries of various eccentricities
and mass ratios, and identify families of {\it stable invariant
loops}. These consist of closed contours in configuration space, such
that as each point initially in them evolves forward in time in the
binary potential, the contour changes its shape but comes back to its
original form and position when the potential has completed a period
(Maciejewski \& Sparke 1997, 2000). When invariant loops do not cross,
they may be filled with gas; if the gas loses or gains angular
momentum it may drift through a sequence of invariant loops.

Although further restrictions may limit the regions occupied by gas in
an eccentric binary, our study offers a firm first survey of the
regions where gaseous disks and planets cannot exist around stars in
eccentric binary systems.

In Section \ref{whatisloop} we review the concept of an {\it invariant
loop}, and describe our orbit integration method and the strategy used
to find invariant loops.  The application to the circular binary as a
test case is presented in Section \ref{ecc0}.  The application to the
general case of arbitrary orbital eccentricity is shown in Section
\ref{eccgeneral}. In Section \ref{observations} we apply this study
to observations of L1551~IRS5 and $\alpha$ Centauri. Conclusions are
presented in section \ref{conclusions}.

\section[]{The Method}\label{method}
\subsection{What is an Invariant Loop?}
\label{whatisloop}

A time-periodic potential in the (two dimensional) orbital plane of an
eccentric binary, is mathematically equivalent to a 3-D system with an
autonomous Hamiltonian, but with the addition of time (in our case,
the binary phase) and the Hamiltonian as two extra dimensions in phase
space (e.g. Lichtenberg \& Lieberman 1992, Section 1.2). If we
restrict ourselves to orbits that lie within the orbital plane of the
binary, the extended phase space will have 6 dimensions and regular
orbits will lie on a 3-dimensional hypersurface. The motion of a
regular orbit is then multiply periodic (Arnold, 1984) with three
frequencies, one of which is the oscillation frequency of the
potential.  If we examine the system at fixed orbital phases, we slice
through the extended phase space at a fixed position on the time axis,
and the resulting projection of a regular orbit will lie on a
2-dimensional hypersurface, densely filling an area on the orbital
plane.

An additional integral of motion would confine an orbit to lie on a
1-dimensional curve. Every time the system comes back to the initial
orbital phase, the particle that follows this orbit will land on the
same 1-dimensional curve: an {\it invariant loop}.  If we were to
``paint'' in a series of different colors all particles that initially
lie on such a curve, we would see each follow its own path as the
binary stars move around each other, distorting the initial shape; but
as the binary returns to the initial orbital phase, the painted curve
would come back to the same initial locus, and although  each particle
would be at a different point along the curve, the original order of
the colors would be conserved.

Stable invariant loops represent the generalization to periodically
time-varying potentials of the stable closed orbits that form the
`backbone' of the orbital structure in time-independent potentials. 
Particle orbits that start near a stable invariant loop will remain
trapped close to it, exploring a nearby region in phase space.  Gas in
a low-viscosity regime trying to settle down in a quasi-static
configuration will converge on the non-intersecting stable invariant
loops. 

\subsection[]{Numerical implementation}\label{numexp}

We write the equations of motion for the binary star system in
term of the eccentric anomaly $\psi$ (e.g. Goldstein 2002, Section
3.7). We use units where the gravitational constant $G$, the binary
semi-major axis $a$, and its total mass $m_1 + m_2$ are set to unity
so that, the binary period is $2 \pi$, and its frequency $\omega =
1$. The separation of the two stars at time $t$, measured from
periastron where the azimuthal angle $\theta=0$, is given by the
radius $r$,

\begin{eqnarray}
r =  a (1-e \cos \psi)\, ,  \label{eq1}\\ 
 \omega t = (\psi - e \sin \psi)  \, ,\\ 
 \cos \theta = a (\cos \psi - e)/r .
\end {eqnarray}

The binary eccentricity, defined as $e=\sqrt{1-b^2/a^2}$
where $a$, and $b$ are the semimajor and semiminor axes,  and the mass
ratio $q = m_2 / (m_1 + m_2)$ are the only free parameters.  We used
an Adams integrator (from the NAG fortran library) to follow the
motion of a test particle moving in the orbital plane of the two
stars.  Kepler's equation (\ref{eq1}) is solved within a tolerance of
$10^{-9}$. In the circular case, the Jacobi energy of the test
particle (computed as a diagnostic for the quality of the numerical
integration) is conserved within one part in $10^9$ per binary period.

The equations of motion of the test particle are solved in an inertial
reference frame using Cartesian coordinates, with their origin at  the
center of mass of the binary.  All test particle trajectories are
launched when the binary is at periastron, with the two components
lying on the $x$-axis.  We search initially for families of loops that
are symmetric about this line, so we launch test particles at various
points along the $x$-axis and perpendicular to it.  We then store the
particle's position and velocity every time the stars return to
periastron.  The computation is halted if the particle runs away,
moving further than 10 times the semimajor axis ($a$) from the center
of mass, or if it comes within a distance of either star that results
in a  high number of force computations, generally due to close
approaches to the stars. In this manner, we obtain orbits around each
star (circumstellar), orbits around both stars (circumbinary), runaway
orbits, and orbits that are captured by one of the stars.

\subsection{Searching for Invariant Loops in a Binary System}\label{seek}

\begin{figure}
\includegraphics[width=84mm]{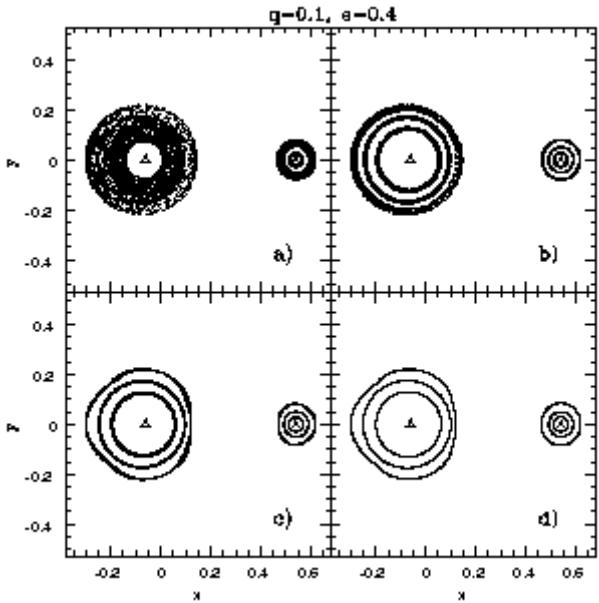}
\vspace{0.1cm}
\caption{Searching for invariant circumstellar loops: successive
positions at periastron of particles launched normally from the
$x$-axis.  In panels a) to d), the launch speeds approach those for
stable loops, and the dispersion of the iterates decreases until 
the loop is found.}
\label{fig.seekloop}
\end{figure}

Among all the possible orbits, we are seeking a very special type: the
{\it invariant loops}, for which the successive phase-space
coordinates of our test particle fall on a one-dimensional curve.  To
find them, we examine the iterates in some two-dimensional subspace,
such as the $x-y$ plane.  We plot the positions of the test particle
at each complete binary period, and compute their dispersion along the
radial direction for those that lie within a sector that spans a small
angle ($5^\circ$) about the x axis. Repeating the integration with the
same starting $x$-value, we adjust the launch velocity $v_y$ to
minimize this dispersion.  In each panel of Figure \ref{fig.seekloop},
we show iterates from three particles that orbit around the primary
star, and three around the secondary.  As $v_y$ is adjusted, the
iterates converge towards the loops in panel d.

Effectively, our method finds only stable loops.  Particles launched
near an unstable loop would rapidly diverge from it, and we would not
see the convergence illustrated in Figure \ref{fig.seekloop}.  Very
near the stars, the stable loops are close to stable circular orbits
in the potential of that star alone.  We exploit this fact to find
families of circumstellar loops: we start by launching a particle
close to either star, with a speed appropriate to the circular orbit.
Once we have found a loop, we move the starting point in small steps
away from the star, and use the loops we have just found to predict
the next starting speed.  The process ends when we can find no more
loops with larger starting distances.  To map out a family of
circumbinary loops, we begin at large radii, where the loops are close
to circular orbits about the center of mass; we continue inwards until
no more loops are found.

\section{Test: Application to the Circular Binary case}\label{ecc0}

In the limit that the binary's eccentricity goes to zero, the
invariant loops should be exactly the closed circumstellar and
circumbinary orbits (periodic orbits) for the circular system. We
computed several orbits for cases with different mass ratios, to 
compare with other theoretical work. 

We took arbitrary points of several computed loops as initial
conditions to calculate the orbit in the non-inertial reference frame,
corotating with the line that joins the stars. The resultant orbits
indeed close on themselves, showing that for the circular binary,
invariant loops are none other than the familiar periodic orbits.

\begin{figure}
\includegraphics[width=84mm]{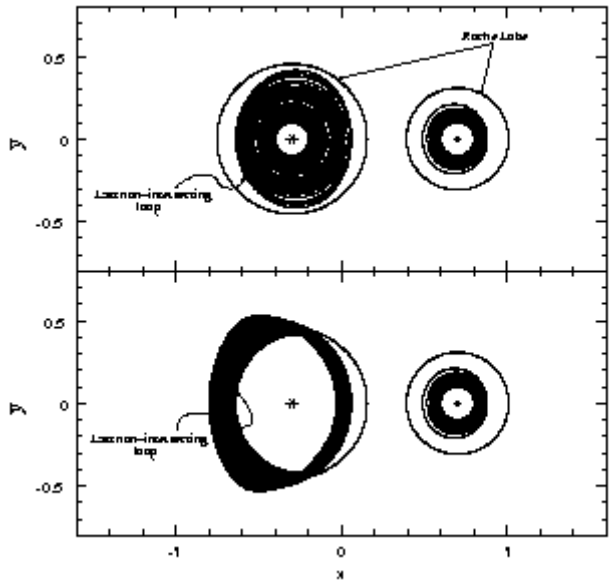}
\vspace{0.1cm}
\caption{Points on the circumstellar invariant loops for the circular
orbit, for a mass ratio q=0.2. Top panel: loops are plotted up to the
position of the last non-intersecting loop (that defines the
circumstellar gaseous disk). Bottom panel: loops are plotted starting
from the last non-intersecting loop until the last loop found.}
\label{fig.periodic}
\end{figure}

In Figure \ref{fig.periodic} we show the invariant circumstellar loops
found for a circular binary. The Jacobi constant is conserved within
$10^{-9}$. In the top panel are plotted the non-intersecting
circumstellar loops, which could be populated by gas particles. In the
bottom panel of this Figure we show an example of the intersecting
loops that start mostly when the disks loops approach to the Roche
lobe. These kind of loops, although stable, can not be populated by
gas particles due to the intersection with the inner disk loops which
would induce shocks dissipating first the less attached gas particles
settle down in the intersecting loops. The dotted circles (aimed with
straight arrows) represent a good fit to the radius of the Roche lobe
as calculated by the analytic approximation by Eggleton (1983).  For
star $i$, the Roche radius $R_i$ is given by

\begin{eqnarray}
\frac{R_i}{a}
\approx  \frac{R_i {\rm (Egg)}}{a}
=\frac{0.49 q_i^{2/3}}{0.6q_i^{2/3}+ln({1+q_i^{1/3}})}
\, ,~ {\rm where} ~~ \\   
q_1=m_1/m_2=\frac{1-q}{q} ~ {\rm and} ~
q_2=m_2/m_1=\frac{q}{1-q} \, .
\label{eq.roche}
\end {eqnarray}

Note that the circumstellar periodic orbits are not circular, but are
elongated perpendicular to the line joining the stars. Orbits very
close to the Roche lobe change their shape dramatically crossing in
some cases this frontier (e.g. circumprimary disk with $q=0.4$) and
producing intersections with the inner orbits.  In an eccentric
binary, the circumstellar disks tend to be also elongated in the same
sense. In high eccentricity cases ($e>0.4$), disks are nearly circular.

\begin{figure}
\includegraphics[width=84mm]{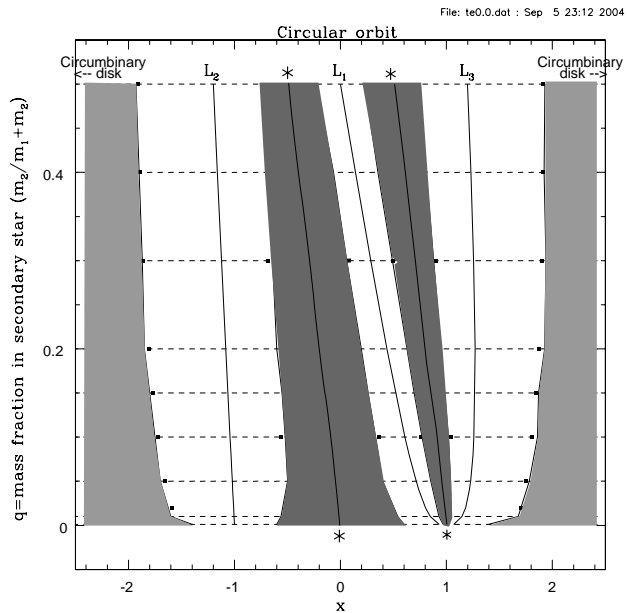}
\vspace{0.1cm}
\caption{Shaded regions show the radii accessible to the circumstellar
and the circumbinary  disks as a function of the mass fraction,
$q={m_2}/({m_1+m_2})$. The Lagrange points $L_1$, $L_2$, and $L_3$ are
indicated, and the position of the stars is shown by near-vertical
lines with a star at the top and bottom of each one. The horizontal
lines show the actual mass fractions where the computations were
done. With filled squares we mark for comparison some examples of the
calculated radii from Rudak \& Paczy\'nski (1981).}
\label{fig.radiuse0}
\end{figure}

Using our method to search for invariant loops, we have computed the
limiting radii of circumstellar and circumbinary disks for mass ratios
$q$ ranging from the Jupiter-Sun system ($q=0.001$) to the equal-mass
case $q=0.5$. The limiting radii are selected as the loops overlap or
when no more loops are found (the more common case when $e>0$). In
Figure \ref{fig.radiuse0}, we present the disk sizes at periastron as
a function of the mass fraction. For our limiting loops, we measured
the disk extent in both directions along the $x$-axis joining the two
stars. The nearly-vertical continuous lines marked at the top with L1,
L2, L3 are the Lagrange points. The circumstellar disks are confined
within the Lagrange points, while the circumbinary disk is
approximately centered on the origin, the mass center of the
binary. The filled squares are the limiting radii found by Rudak \&
Paczy\'nski (1981), who approximate the gas streamlines in
pressureless accretion disks by non-intersecting simple periodic
orbits. They compute in this way several disk sizes allowed by this
potential. The results of our computations for disk sizes in the
circular case are in very good  agreement.

\section{The General Case: Eccentric binaries}\label{eccgeneral}

Unlike the zero eccentricity case, where particles see a simple
time-steady potential, we now have a variation of the potential in
time;  in this manner particles will see a changing situation in any
rotating reference frame. The phase space increases its dimensionality
including now time as a canonical variable.

\begin{figure}
\includegraphics[width=84mm]{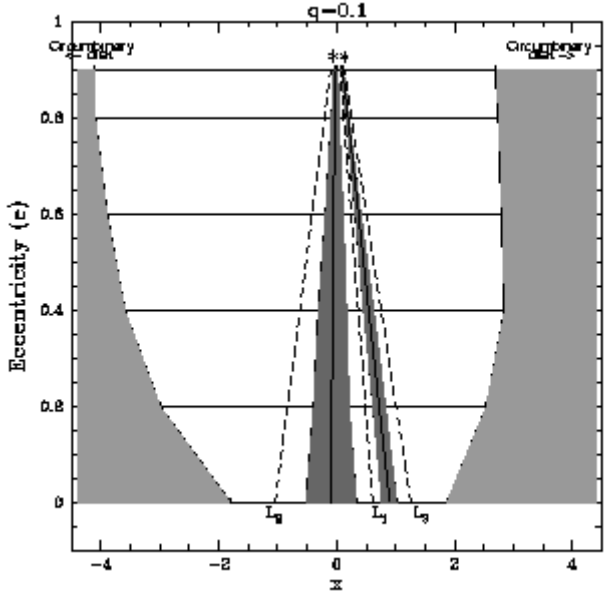}
\vspace{0.1cm}
\caption{Similar to figure \ref{fig.radiuse0} (disks radii are
computed at periastron) but with orbital eccentricity as ordinate for
the case q=0.1.  The position of the stars is shown by near-vertical
lines with a star at the top of each one. The horizontal  lines show
the actual eccentricities where the computations were done.  The
dashed lines marked as $L_1$, $L_2$, and $L_3$ are the Lagrangian
points  for the circular binary, multiplied by $(1-e)$.}
\label{fig.q0.1}
\end{figure}

In the same manner as for the circular case, we have constructed
figures showing disk sizes for eccentricities up to $e=0.9$, for a
given mass ratio $q$. In Figure \ref{fig.q0.1} we show the regions
accessible to circumstellar or circumbinary disks for a binary with a
fixed mass ratio of $q=0.1$ as the eccentricity grows. $L_1$, $L_2$,
and $L_3$ are the Lagrangian points in the pericenter of the system
multiplied by $(1-e)$. The circumstellar disks become rapidly smaller
as the binary grows more eccentric, while the circumstellar disk
recedes. Additionally, in Table \ref{avradS1}, we present the averaged
radii of the circumstellar disks. In Table \ref{diskcompS2L} we
compare the averaged radius of the circumsecondary and circumprimary
disks, respectively, with the Lagrangian radius computed at the
pericenter multiplied by $(1-e)$.  For all mass ratios and $e\leq
0.8$, the size of the disk around the secondary star is well predicted
as $0.4 \pm 0.03$ of the Lagrange radius times $(1-e)$.

We have fitted a power law in both $e$ and $q$ 
to the calculated
radii, 
for eccentricities in the  range [0.00, 0.9], and $q$ in the range 
[0.01, 0.99], obtaining the following 
relation for the size of the circumstellar disks: 

\begin{equation}
R_i \approx R_{i, {\rm Egg}} \times 0.733\ (1-e)^{1.20}\ q^{0.07}\ R_2 \, ,
\label{eq.Rfloop}
\end {equation}

where $ R_{i, {\rm Egg}}$ is Eggleton's estimate for the Roche Lobe
average radius  (equation \ref{eq.roche}). In Table \ref{diskcompS2LR}
we have calculated the ratio between this quantity and the actual
radii of the circumstellar disks computed; our formula is  accurate to
$\pm 6.5\%$.

\begin{figure}
\includegraphics[width=84mm]{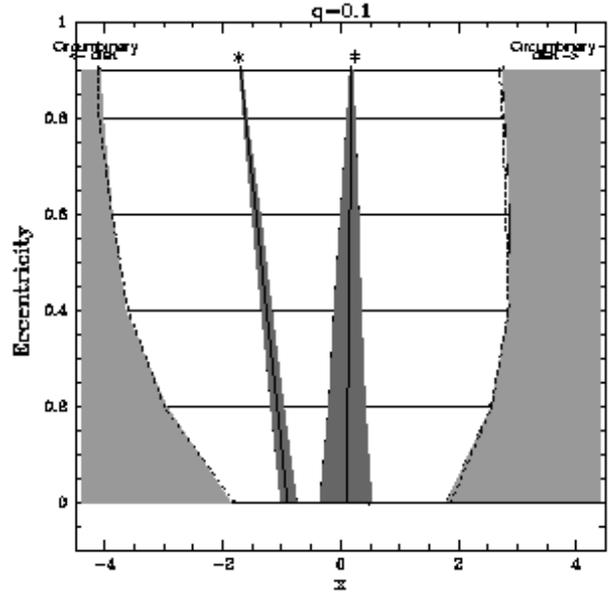}
\vspace{0.1cm}
\caption{As Figure \ref{fig.q0.1}, but at apastron: boundaries of the
circumstellar disks and the inner edge of  the circumbinary disk, for
mass ratio $q=0.1$, as a function of eccentricity $e$. Dashed lines
indicate the inner boundary of the circumbinary disk in Figure
\ref{fig.q0.1}: showing that the position of this disk changes little
with binary phase.}
\label{fig.q0.1_apo}
\end{figure}

\begin{figure}
\includegraphics[width=84mm]{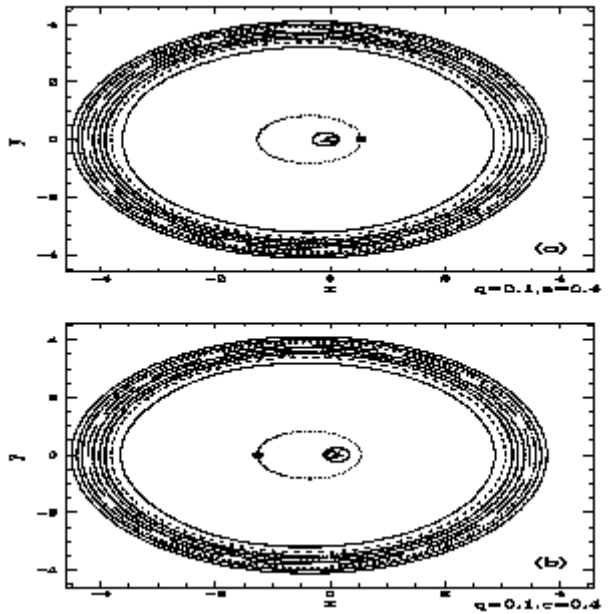}
\vspace{0.1cm}
\caption{Circumbinary disk calculated at the binary pericenter (a), 
and at the apocenter (b), for the case $q=0.1$, $e=0.4$. Dotted 
curves in the center show the stars trajectories; and small curves
surrounding the stars (which are pointed by triangles) show the 
extent of circumstellar disks.}
\label{fig.CBperiapo}
\end{figure}

In Table \ref{lrCB} we show the left and right radii of the
circumbinary disk measured from the center of mass. A change in the
mass fraction leaves the size and shape of the circumbinary disk
almost unaffected. On the other hand, a slight increase in the
eccentricity will result in a noticeable change as can be appreciated
from Figure \ref{fig.q0.1}; The circumbinary disk is now off-center
with respect to the center of mass; it almost preserves a constant
distance from the closest approach of either star: see figures
\ref{fig.q0.1_apo} and \ref{fig.CBperiapo}.

We have calculated the approximate position of the closest resonances
at the edge of the circumstellar and circumbinary disks in all the
computed cases presented. For the circular case, for mass ratios
$q\geq 0.01$ all disks finish approximately at their 1:3
resonance. For an extreme mass ratio, as with  the Sun-Jupiter case,
the circumbinary and circumprimary disks can extend further, to the
1:2 resonance. As the eccentricity increases, the disk is truncated at
successively higher order resonances.  For $e\approx0.2$ the closest
resonances to the end of the circumstellar disks are 1:5 or 1:6. For a
higher eccentricity ($e>0.6$) the closest resonances are of even
higher order (1:8 to 1:20). For the circumbinary disk at
eccentricities, $e\geq 0.1$, the closest resonances to the inner
boundary are 1:4 or 1:5.

\begin{table*}
\centering
\begin{minipage}{140mm}
\caption{Averaged radius of disk around star 2 (secondary for 
$q<0.5$, primary for $q>0.5$) in units of the semimajor axis, $a$. All radii are measured from the outermost loop at the binary periastron.}
\begin{tabular}{@{}llrrrrlrlr@{}}
\hline
$m_2$ / $e$     & 0.0  & 0.2 & 0.4 & 0.6 & 0.8 \\ 
\hline
0.1  &0.125  &0.100  &0.079  &0.049  &0.019  \\
0.2  &0.162  &0.130  &0.098  &0.048  &0.029  \\
0.3  &0.195  &0.165  &0.097  &0.067  &0.028  \\
0.4  &0.228  &0.195  &0.125  &0.083  &0.033  \\
0.5  &0.257  &0.213  &0.147  &0.097  &0.037  \\
0.6  &0.317  &0.228  &0.153  &0.093  &0.047  \\
0.7  &0.350  &0.225  &0.171  &0.109  &0.037  \\
0.8  &0.387  &0.260  &0.187  &0.126  &0.049  \\
0.9  &0.426  &0.297  &0.231  &0.141  &0.064  \\
\hline
\end{tabular}
\label{avradS1}

\caption{Averaged radius of the disk around star 2 (secondary for
$q<0.5$, primary for $q>0.5$) compared  with Lagrangian radii
calculated at the pericenter: $\frac{1}{2}(L1-L)(1-e)$, where $L=L3$
for  $q=[0.1,0.5]$, and $L=L2$ for $q=[0.5,0.9]$.}
\begin{tabular}{@{}llrrrrlrlr@{}}
\hline
$m_2$ / $e$ & 0.0  & 0.2 & 0.4 & 0.6 & 0.8 \\
\hline
0.1  &0.38  &0.39  &0.40  &0.38  &0.30  \\
0.2  &0.39  &0.39  &0.39  &0.29  &0.34  \\
0.3  &0.40  &0.42  &0.33  &0.34  &0.29  \\
0.4  &0.42  &0.44  &0.38  &0.38  &0.30  \\
0.5  &0.43  &0.44  &0.41  &0.40  &0.31  \\
0.6  &0.49  &0.44  &0.39  &0.36  &0.36  \\
0.7  &0.50  &0.40  &0.40  &0.39  &0.26  \\
0.8  &0.51  &0.43  &0.41  &0.41  &0.33  \\
0.9  &0.52  &0.45  &0.47  &0.43  &0.39  \\
\hline
\end{tabular}
\label{diskcompS2L}

\caption{Same as Table \ref{diskcompS2L} but the disk radius is
compared  with the radius $R_i$ obtained from equation
\ref{eq.Rfloop}.}
\begin{tabular}{@{}llrrrrlrlr@{}}
\hline
$m_2$ / $e$ & 0.0  & 0.2 & 0.4 & 0.6 & 0.8 \\
\hline
0.1  &0.93  &0.98  &1.09  &1.10  &0.98  \\
0.2  &0.91  &0.95  &1.02  &0.81  &1.12  \\
0.3  &0.92  &1.01  &0.84  &0.94  &0.91  \\
0.4  &0.93  &1.04  &0.94  &1.02  &0.93  \\
0.5  &0.93  &1.00  &0.98  &1.05  &0.92  \\
0.6  &1.01  &0.95  &0.90  &0.89  &1.04  \\
0.7  &0.99  &0.83  &0.89  &0.93  &0.72  \\
0.8  &0.96  &0.84  &0.85  &0.93  &0.83  \\
0.9  &0.87  &0.80  &0.88  &0.87  &0.91  \\
\hline
\end{tabular}
\label{diskcompS2LR}

\caption{Coordinates where the inner edge of the  circumbinary disk
crosses the $x$ axis, in units of the semimajor axis, $a$}
\begin{tabular}{@{}llrrrrlrlr@{}}
\hline
$m_2$ / $e$     & 0.0  & 0.2 & 0.4 & 0.6 & 0.8 \\
\hline
0.1  &-1.80,1.87   &-3.00,2.54   &-3.60,2.85   &-3.90,2.80   &-4.10,2.75 \\
0.2  &-2.00,2.04   &-3.00,2.56   &-3.70,3.05   &-3.90,2.86   &-4.00,2.84 \\
0.3  &-1.90,1.94   &-3.20,2.98   &-3.60,3.12   &-3.80,3.16   &-4.00,3.33 \\
0.4  &-1.90,1.92   &-3.10,2.98   &-3.50,3.27   &-3.70,3.40   &-3.70,3.30 \\
0.5  &-2.00,2.00   &-2.70,2.70   &-2.90,2.90   &-3.40,3.40   &-3.40,3.40 \\
\hline
\end{tabular}
\label{lrCB}

\end{minipage}
\end{table*}

\section{Morphology and Temporal evolution}
\label{disksevol}

\begin{figure}
\includegraphics[width=84mm]{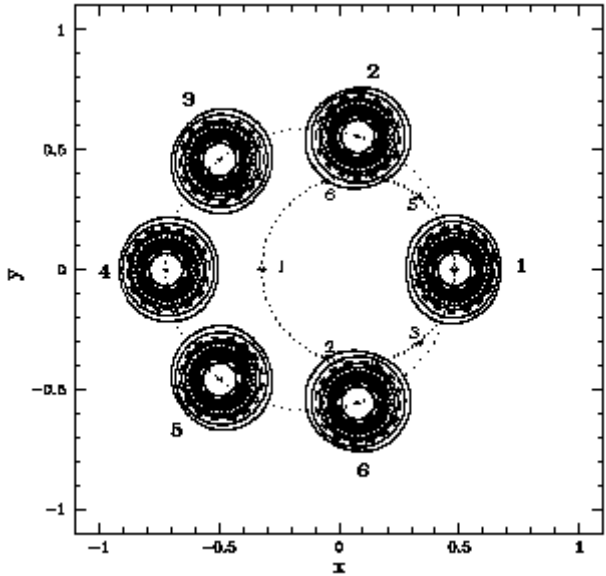}
\vspace{0.1cm}
\caption{The circumsecondary disk at 6 different phases of the binary,
for the case $q=0.4$, $e=0.2$. Dotted lines show the orbits of both
primary and secondary stars; asterisks show the positions at
corresponding times.}
\label{fig.diskevol}
\end{figure}

Although the disks were calculated  when the stars are at their
pericenter, they can be followed to any other phases of the binary. In
Figure \ref{fig.diskevol}, we see that the morphology of the disks
changes slightly for different phases.  The maximum change in radius
is generally about $5\%$ of the radius measured at pericenter,
increasing or decreasing  with no preference for any specific phase.

The loops that form the circumstellar disks are  nearly circular close
to the corresponding star, up to about 80\% of the  outer radius;
there  the shape can change abruptly to become more elliptical. This
is more pronounced when the binary is {\it less}  eccentric, because
the disk then  extends further in radius.  The loops are also slightly
off-centered.  We can measure this by sampling the distance $s$ of
each circumstellar loop from the star at $N$ points equally spaced in
azimuthal angle $\phi$ around it, and computing the Fourier
coefficients

\begin{eqnarray}
A_k=\frac{1}{N} \sum_{i=0}^{N} s(\phi) cos(k\phi),  \,
\label{LopEll}\\ 
B_k=\frac{1}{N} \sum_{i=0}^{N} s(\phi) sin(k\phi) \, .
\end {eqnarray}

The lopsidedness of the disks at the binary pericenter is measured by
computing $\sqrt{A_1^2+B_1^2}/r_{peri}$, (and the ellipticity by
$\sqrt{A_2^2+B_2^2}/r_{peri}$), where $r_{peri}$  is the
azimuthally-averaged radius of the circumstellar loop.  We find that
the maximum lopsidedness of the outermost circumstellar loops at
pericenter is $5\%$.  The lopsidedness changes by less than  $1\%$
with the binary phase.

For the circumstellar disks we measure the maximum ellipticity at
pericenter by computing  $ell=1-b/a$, the ratio of the major  axis,
which  is always perpendicular to the line that joins the stars, to
the minor  axis which lies along the line joining the stars.  The
ellipticity of the outermost circumstellar loop at the pericenter is
higher in the binaries with lower eccentricities ($e<0.4$), since
these  disks extend much further  inward in radius (see Figure 5); it
is generally in  the range $0.08 < ell < 0.2$.   For the circumbinary
disks the maximum ellipticity of the inner edge is generally reached
when the stars are at the apocenter and is $ell\approx0.05$. The
maximum change in the ellipticity for the circumbinary disks with
phase is $5\%$ of the ellipticity at the pericenter.

\section{Application to Observations}\label{observations}

Observational parameters of only a very few binaries with accretion or
proto-planetary disks are known. We have chosen  as an application a
couple of well-known examples of these kind of systems for which some
important parameters (like the size of the observed disks or the
orbital parameters) are known; the first is related to accretion
(L1551), and the second to protoplanetary disks ($\alpha$ Centauri).

In this case, for example, if the disks radii and the distance between
the stars are known, we can approach the orbital parameters and mass
ratios. On the other hand, if the orbital parameters, like
eccentricity, and semimajor axis of the binary orbit are known, the
potential sizes of the circumstellar and circumbinary disks can be
readily calculated.

\subsection{Circumstellar disks around the L1551 binary}\label{L1551}

\begin{figure}
\includegraphics[width=84mm]{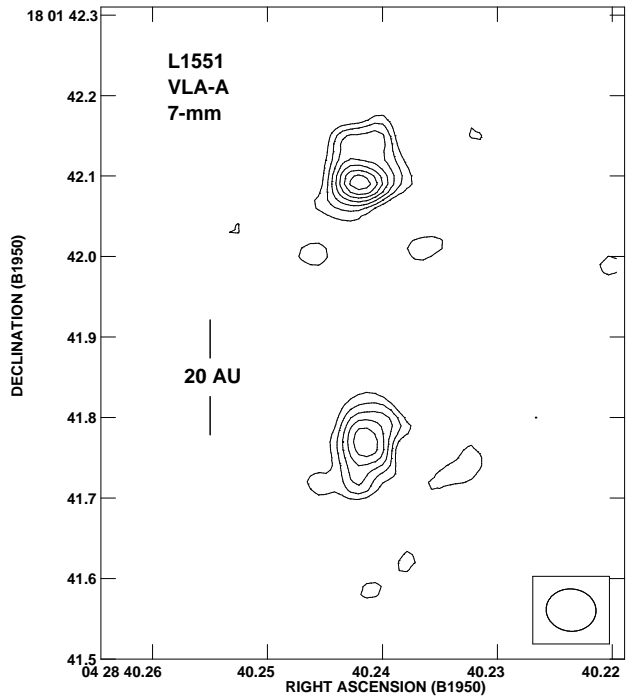}
\vspace{0.1cm}
\caption{Cleaned, natural-weight VLA map of the L1551~IRS5 region at
7-mm. Figure 1 of Rodr\'\i guez et al. (1998). The emission shown in 
this map is interpreted to arise from two compact protoplanetary structures
forming a gravitationally-bound binary system.}
\label{fig.l1551}
\end{figure}

Rodr\'{\i}guez et al. (1998, 2003) report interferometric observations
at 7~mm showing hot dust in the core of the star-forming  region
L1551, a molecular cloud in Taurus. In the resolved core of L1551 are
two distinct disks with a separation of 45~$AU$ that appear to be
circumstellar disks associated with a binary system (Figure
\ref{fig.l1551}). With accretion disk models they find that best fits
to the data are obtained for disks with semi-major axes of about
10~$AU$ and total masses of approximately 0.05~M$_\odot$, enough to
form protoplanetary disks. The disks are apparently elongated towards
each other. Since the circumstellar loops tend to be elongated in the
perpendicular sense, this suggests that the binary orbit is tilted an
angle of approximately $60^{\circ}$  to our line of sight, as models
of the system and its jets indicate (Rodriguez et al. 2003; Osorio et
al. 2003). The stars now lie close to the line of nodes of the orbit
so the observed separation is close to the true separation
(Rodr\'{\i}guez et al. 2003). Millimetre-wave observations suggest the
system is embedded in an elongated structure of dust and gas with
scales of 100-400 $AU$ (Lay et al. 1994; Keene \& Mason 1990). This
structure may correspond to a circumbinary ring or disk (Looney et
al. 1997), that would probably be providing enough material to fill
the circumstellar disks.

\begin{table*}
\centering
\begin{minipage}{140mm}
\caption{Binary parameters for possible L1551~IRS5 models, and the 
implied sizes of the circumstellar disks}
\begin{tabular}{@{}llrrrr@{}}
\hline
     & present  &          &           &            &  \\
$q, e$     & phase    & $a\ (AU)$ & $r_{sec}$ & $r_{prim}$ & $r_{CB}$ \\
\hline
$q=0.5, e=0.0$ &      & 45   & 13.5  & 13.5 & 88 \\
$q=0.4, e=0.0$ &      & 45   & 9.5   & 15.3 & 86 \\
$q=0.5, e=0.2$ & peri & 56   & 11.2  & 11.2 & 152 \\
$q=0.5, e=0.2$ & apo  & 37.5 & 7.5   & 7.5  & 101 \\
$q=0.4, e=0.2$ & peri & 56   & 14.6  & 7.3  & 157 \\
$q=0.4, e=0.2$ & apo  & 37.5 & 9.75  & 4.9  & 105 \\
\hline
\label{tab.L1551}
\end{tabular}
\end{minipage}
\end{table*}

Proper motions do not give clear information about the orbit, but we
can use the disk sizes to set bounds on the eccentricity and the mass
ratio of the system.  For any fixed eccentricity, Table
\ref{tab.L1551} shows that for a mass ratio $q=0.4$, or $m_1:m_2 =
3:2$, the circumsecondary disk should be roughly 60\% smaller than the
circumprimary disk. Since we see two nearly equal disks, this implies
$0.4 \leq q \leq 0.5$ for this system. If we assume that is a circular
orbit, the circumstellar disks could both extend to 13.5 $AU$ if the
stellar masses are equal, or 9.5 and 15.3 $AU$ for circumsecondary and
circumprimary disks respectively, if $q=0.4$.  The circumbinary disk
gap would then end about 90 $AU$ from the center of mass of the
system.  The measured disks are only approximately 10 $AU$ in radius,
which might be probably due simply to the fact that the emission
weakens, make it difficult to detect, however we have also explored
the possibility that the reduced size of the disks is due to the fact
that the binary orbit is slightly eccentric, in this manner we have
constructed a range of possibilities for this system. For $e>0.2$,
Table \ref{tab.L1551} shows that the disks should be truncated short
of their observed extent of 10~$AU$; so we conclude that the
eccentricity of the system is  in the interval $e=[0, 0.2]$. The
extreme case with $q=0.4$, and $e=0.2$, would produce very different
disk sizes ($r_{prim}/r_{sec} \sim 1/2$), so we discard this
possibility also. Thus the system parameters are constrained
approximately by $(0.5-q) + 0.5 e \leq 0.1$.

\subsection[]{Zones for Planets around Alpha Centauri}\label{acentauri}

As one of the alternatives to form planets in accretion disks arised
the ``planetesimal theory'': planets are formed in circumstellar dust
disks, where colliding dust grains accrete into planetesimals
approximately of 1-10 km of diameter (Safronov 1969; Lissauer 1993).
As the sizes of particles increase and gravitational attraction
becomes more important, the cross-section for accretion also
increases, as more and more particles collide.  This effect, combined
with dynamical friction, leads to ``runaway growth'': large bodies
accrete more efficiently than small bodies, becoming planetary embryos
that increase their masses through collisions until most of the
material is accreted or dispersed by these new protoplanets (Greenberg
et al. 1978; Wetherhill \& Stuart 1989). Assuming that planets are
initally formed in accretion disks, invariant loops can give us also a
good idea of what possibilities a binary system has to harbor stable
planets, and the maximum radius of a protoplanetary disk that can give
rise to them.

$\alpha$ Centauri is composed of a binary star consisting on a G2 star
with 1.1 M$_\odot$ ($\alpha$ Centauri A) and a K1 star with 0.91
M$_\odot$ ($\alpha$ Centauri B) and a third component ($\alpha$
Centauri C or Proxima Centauri), which is thought to orbit this pair
but at a very large distance (12,000 $AU$). The binary system has an
eccentricity of 0.52 and a semimajor axis of 23.4 $AU$ (See 1893, 
Heintz 1982).

This system and the stability of planets around them has been
extensively studied. Wiegert \& Holman (1997) used direct numerical
integration to find that planets orbiting prograde in the plane of
$\alpha$ Centauri A and B within 3~$AU$ of either star can be stable for
several million years, as can planets in circumbinary orbit more than
70~$AU$ from the center of mass of the system.  Marziari \& Scholl
(2000) studied the evolution of planetesimals perturbed by gas drag in
a disk in the plane of the binary system, finding that planetesimals
are able to accrete one another within 2 $AU$ of $\alpha$ Cen
A. Quintana et al. (2002) follow the growth of planetary embryos under
the gravitational forces of the binary system using a symplectic
N-body accretion algorithm, finding stable planets in prograde orbits
up to 2.5~$AU$ from either star, and showing how for high inclinations of
the orbits, with respect to the plane of the binary, they become 
unstable.
 
\begin{figure}
\includegraphics[width=84mm]{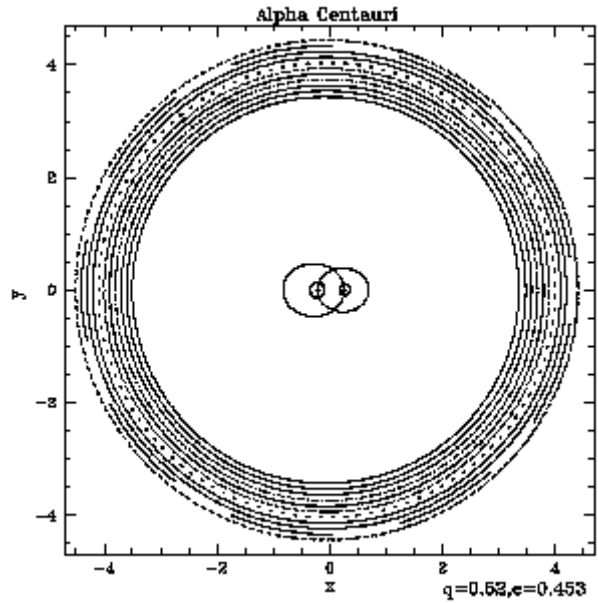}
\vspace{0.1cm}
\caption{Circumbinary disk constructed with invariant loops applied to
the case of $\alpha$ Centauri. Ellipses at the center mark the stellar
trajectories.  Smaller curves around the two three-pointed stars mark
the limit of  the circumstellar disks.}
\label{fig.totacen}
\end{figure}

\begin{figure}
\includegraphics[width=84mm]{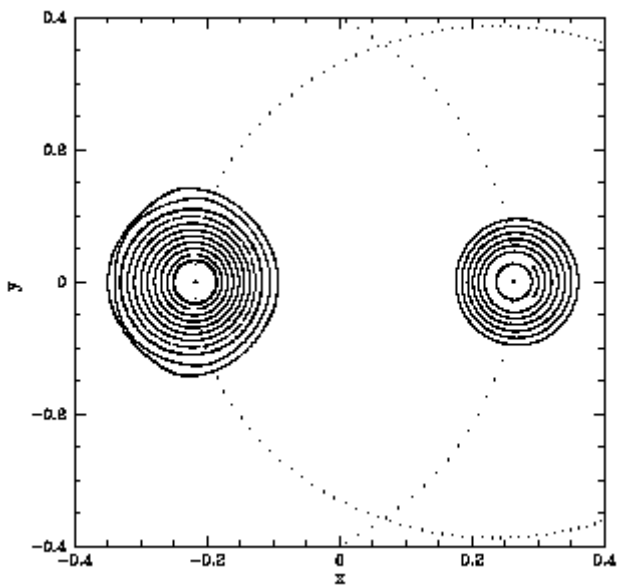}
\vspace{0.1cm}
\caption{A zoom of the circumstellar disks shown in Figure
\ref{fig.totacen}.}
\label{fig.zoomacen}
\end{figure}

\begin{figure}
\includegraphics[width=84mm]{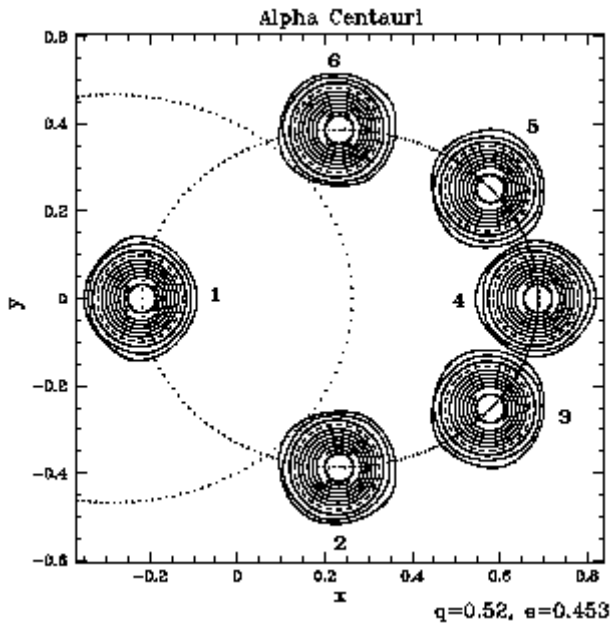}
\vspace{0.1cm}
\caption{Six phases of the $\alpha$ Centauri circumprimary disk 
constructed with invariant loops. Dotted curves show stellar trajectories.
Numbers show the ordered phases, as in Figure \ref{fig.diskevol}.}
\label{fig.evolacen}
\end{figure}

Circumbinary and circumstellar disks constructed with invariant loops
applied to the case of $\alpha$ Centauri are presented in Figures
\ref{fig.totacen}  and \ref{fig.zoomacen}.   We calculated a size of
3~$AU$ for the circumstellar disk (or potential protoplanetary disk)
around $\alpha$ Centauri A, and 2.3~$AU$  for $\alpha$ Centauri B.  The
circumbinary gap, measured from the center of mass, should extend to
80~$AU$.

In Figure \ref{fig.evolacen} we show the evolution of the
circumprimary disk ($\alpha$ Centauri A) with the phase.  The disk
changes its shape and slightly its size as it  evolves in time. While
it is unlikely that such small changes  in shape could be observed
directly,  the crowding of orbits at phases 1, 2, and 5 are likely to
lead to increased heating and inflow when the disk contains gas.

\section{Conclusions}\label{conclusions}

With the concept of {\it invariant loop}, we have extended the
possibilities for orbital studies in accretion or protoplanetary disks
of binary stellar systems, with no restriction on the mass ratio ($q$)
or eccentricity ($e$) of the binary. In this manner the method
represents an extension to the periodic-orbit analysis in the
well-known circular case. Sweeping the parameter space, we are able to
address the limits and possibilities for gaseous circumstellar and
circumbinry disks from the point of view of the ``pure'' potential
exerted by the binary. In the case of accretion disks, this
approximation would closely represent the low viscosity regime of the
gas. Although the studies presented in this work were restricted to
the plane of the disks, this technique can also be applied to orbits
out of the plane of the binary orbit. Compared with high-resolution
hydrodynamic simulations the method is computationally cheap and fast.

We find that the size of the circumstellar and circumbinary disks
depends strongly on the eccentricity.  The average radius of the
circumstellar disks is approximately 40$\%$ of the Lagrangian radius
calculated at the pericenter of the binary orbit. Equation
\ref{eq.Rfloop} provides a more accurate analytic approximation to the
disk size.

The inner radius of the circumbinary disks  (the {\it gap} radius) is
practically independent of the mass ratio. For the circular binary,
the circumbinary disk ends close to the 1:3 resonance for non-extreme
mass ratios ($q\geq 0.01$). For higher eccentricities, the inner edges
of the circumbinary disks appear to be truncated close to higher order
resonances (1:4, to 1:6).  Even a slight increase in eccentricity for
a given mass ratio  will cause the gap to grow to much larger
radii. The inner edge of a circumbinary disk is not centered at the
center of mass; instead particles in the disks appear to pass at
near-equal distances from the apocenter positions of the two stars.

In the temporal evolution, 
the circumbinary disks show almost no change with binary phase, while
the circumstellar disks change their shape and size only
slightly. The average radius changes by only $5\%$ at different
phases 
of the binary.  The largest or smallest radius can be reached at any phase,
i.e., there is no preference for larger radii at any specific phase.

Near the outer edge, the shape of the disks become less circular in
general and slightly off-centered, by up to 5\% The change of
lopsidedness with the phase is less than $1\%$ of  the lopsidedness
measured at the pericenter of the binary.

For the circumstellar disks the maximum ellipticity is in most cases
reached at pericenter, where the semimajor axes of the circumstellar
disks are perpendicular to the line that joins the stars. The
ellipticity at the pericenter is higher in the cases with lower
eccentricities  ($e<0.4$)and it is generally below $ell \approx 0.2$.
For larger eccentricities the ellipticity of the circumstellar disks
is $ell<0.08$. The maximum change of ellipticity with time is $7\%$ of
the ellipticity measured at the pericenter of the binaries.

For the circumbinary disks the maximum ellipticity is generally reached 
when the stars are at the  apocenter and is $ell\approx0.05$. The maximum 
change in the ellipticity for the circumbinary disks with phase is $5\%$ 
of the ellipticity at the pericenter.

As an application to observations we selected two specific objects:
L1551-IR5 and $\alpha$ Centauri. Assuming that the line that joins the
stars rests approximately in the plane of the sky, the eccentricity of
the system in L1551-IR5 is $e\leq 0.2$ and the mass ratio is $q\approx
0.5$. The circumbinary inner edge disk should lie approximately at
90~$AU$ if $e=0$, and in the interval 100-150 $AU$ for the extreme
case $e=0.2$. In the case of $\alpha$ Centauri, we calculated the
zones for planets around each star and around the binary. We find that
for $\alpha$ Cen A, the last stable loop reaches 3 $AU$, while for
$\alpha$ Cen B it is at 2.3 $AU$. Beyond the gap, the innermost
invariant loops start at a distance of 80 $AU$ from the center of mass.

Every day, disks are being discovered in more binary (and multiple)
stellar systems.  Theoretical studies of these systems can help to
restrict their geometrical characteristics, and may also help to
constrain some other unknown physical characteristics like viscosity.

\section*{Acknowledgments}
We thank Luis Felipe Rodr\'\i guez for supplying Figure
\ref{fig.l1551}. L. S. and B. P. thank NASA 
and the Space Telescope Science Institute for support 
through grant HST-AR-09522.01-A. 
L. A. is grateful to the Department of Astronomy
of the University of Wisconsin-Madison for a fruitful
sabbatical stay where this work got started. He also acknowledges
support from DGAPA/UNAM through grant IN113403.  B. P. thanks Mexico's
CONACyT for a postdoctoral fellowship, and NASA for support through 
grant NAG5-10823.

\label{lastpage}

\end{document}